\documentclass{vldb}
\usepackage{graphicx}
\usepackage{array}
\usepackage{caption} 
\usepackage{float}
\usepackage{balance}  
\usepackage{units}   
\usepackage{verbatim}
\usepackage{todonotes}
\usepackage[square,numbers,sort]{natbib}     
\usepackage[hyphens]{url}     
\usepackage{xspace}     
\usepackage{multirow}
\usepackage{graphicx}     
\usepackage[labelformat=empty,justification=centering]{subfig}
\usepackage{booktabs}     
\usepackage[vlined,linesnumbered,ruled,noend]{algorithm2e}  
\usepackage{amsmath,amssymb,amsfonts}
\usepackage{siunitx}

\newcommand{\spara}[1]{\smallskip\noindent\textbf{#1}}
\newcommand{\mpara}[1]{\medskip\noindent\textbf{#1}}

\newenvironment {squishlist}
{\begin{list}{$\bullet$}
  { \setlength{\itemsep}{0pt}
     \setlength{\parsep}{3pt}
     \setlength{\topsep}{3pt}
     \setlength{\partopsep}{0pt}
     \setlength{\leftmargin}{1.5em}
     \setlength{\labelwidth}{1em}
     \setlength{\labelsep}{0.5em} } }
{\end{list}}

\newcommand{\testset}{\ensuremath{E_{TS}}\xspace}
\newcommand{\trainset}{\ensuremath{E_{TR}}\xspace}
\newcommand{\inse}{\texttt{INS}\xspace}
\newcommand{\rmve}{\texttt{RMV}\xspace}

\vldbTitle{Link Prediction via Higher-Order Motif Features}
\vldbAuthors{Abuoda, De Francisci Morales, and Aboulnaga}
\vldbDOI{https://doi.org/TBD}
\vldbVolume{12}
\vldbNumber{xxx}
\vldbYear{2019}

\begin{document}


\title{Link Prediction via Higher-Order Motif Features}

\numberofauthors{8} 

\author{
%
%
\alignauthor
Ghadeer Abuoda\\
       \affaddr{College of Science and Engineering, HBKU}\\
       \affaddr{Doha, Qatar}\\
       \email{gabuoda@hbku.edu.qa}
\alignauthor
Gianmarco De~Francisci~Morales\\
       \affaddr{ISI Foundation}\\
       \affaddr{Turin, Italy}\\
       \email{gdfm@acm.org}
\alignauthor
Ashraf Aboulnaga\\
       \affaddr{Qatar Computing Research Institute, HBKU}\\
       \affaddr{Doha, Qatar}\\
       \email{aaboulnaga@hbku.edu.qa}
}


\maketitle

\begin{abstract}

Link prediction requires predicting which new links are likely to appear in a graph.
Being able to predict unseen links with good accuracy has important applications in several domains such as social media, security, transportation, and recommendation systems. 
A common approach is to use features based on the common neighbors of an unconnected pair of nodes to predict whether the pair will form a link in the future.
In this paper, we present an approach for link prediction that relies on higher-order analysis of the graph topology, well beyond common neighbors.

We treat the link prediction problem as a supervised classification problem,
and we propose a set of features that depend on the patterns or \emph{motifs} that a pair of nodes occurs in.
By using motifs of sizes 3, 4, and 5, our approach captures a high level of detail about the graph topology within the neighborhood of the pair of nodes, which leads to a higher classification accuracy.
In addition to proposing the use of motif-based features, we also propose two optimizations related to constructing the classification dataset from the graph.
First, to ensure that positive and negative examples are treated equally when extracting features,
we propose adding the negative examples to the graph as an alternative to the common approach of removing the positive ones.
Second, we show that it is important to control for the shortest-path distance when sampling pairs of nodes to form negative examples, since the difficulty of prediction varies with the shortest-path distance.
We experimentally demonstrate that using off-the-shelf classifiers with a well constructed classification dataset results in up to 10 percentage points increase in accuracy over prior topology-based and feature learning methods.

\end{abstract}


\section{Introduction}
\label{sec:intro}
Given a graph $G(V,E)$ at time $t_1$, the \emph{link prediction} problem requires finding which edges $\{e \not\in E\}$ will appear in the graph at time $t_2 > t_1$\protect\cite{liben2007link}.
Predicting which new connections are likely to be formed is a fundamental primitive in graph mining, with applications in several domains.
In social media, friend and content recommendations are often modeled as link prediction problems\protect\cite{aiello2012friendship}.
Link prediction has also been used to detect credit cart fraud in the cybersecurity domain\protect\cite{wang2015link,liben2007link},
to predict protein-protein interactions in bioinformatics~\cite{airoldi2006mixed},
for shopping and movie recommendation in e-commerce~\cite{chen2005link},
and even to identify criminals and hidden groups of terrorists based on their activities~\cite{al2006link}.

Traditionally, link prediction models rely on topological features of the graph, and on domain-specific attributes of the nodes (usually to induce a similarity function)~\cite{al2011survey}.
Most topological features are based on common neighbors, i.e., they rely on the idea of `closing triangles'~\cite{newman2001clustering}.
More advanced approaches such as non-negative matrix factorization (NMF) and graph embeddings have also been tried recently~\cite{menon2011link, tsitsulin2018verse}.
However, traditional topological features that rely on common neighbors, such as the Jaccard index and Adamic/Adar measure~\cite{adamic2003friends}, have proven to be very strong baselines which are hard to beat~\cite{tsitsulin2018verse}.

These traditional features are not only effective, but also efficient to compute as they originate from triadic graph substructures.
Fortunately, recent developments in algorithms and systems have improved our ability to efficiently count motifs with more than three nodes~\cite{ahmed2015efficient,teixeira2015arabesque}.
Therefore, given the outstanding results of traditional topological features, it is natural to wonder about the predictive power of more complex features based on these higher-order motifs.
In this paper, we show that higher-order motif features significantly improve the accuracy of link prediction models.

The present work focuses only on topological features, as node attribute features are domain- and application-specific, and are orthogonal in scope.
As is common practice, we cast the link prediction problem as a binary classification task.
We train a machine learning model on a sample of node pairs from the graph,
where pairs with an edge between them represent a positive example, and pairs without an edge represent a negative one~\cite{fire2011link}.

While developing our methodology, we faced several technical issues in extracting features for the link prediction problem that have not been explicitly addressed in the literature so far.
Two issues deserve particular attention: how to generate motif features in a way that is consistent between training and testing, and how to select the negative examples for the dataset.
For the first issue, the common practice is to remove a set of existing edges from the graph (the positive test set), and then train the classifier on the remaining edges.
Here we propose an alternative based on adding a set of negative examples (non-existing edges) to the graph when extracting the features.
This version of the features consistently outperforms the former in terms of accuracy.
For the second issue, we study the effect of node distance for negative examples on classification accuracy.
We show that distance is an important factor that should be controlled for when creating a dataset (an under-appreciated fact in the link prediction literature~\citep{yang2015evaluating}).
This effect is consistent with the phenomenon of locality of link formation in growing networks~\citep{leskovec2008microscopic}.

A shorter version of this study appears in~\cite{ecml19linkpred}. Our main contributions are as follows:
\begin{squishlist}
\item We show that more complex topological features based on higher-order motifs are powerful indicators for the link prediction problem in a variety of domains;
\item These features improve the accuracy of standard classifiers by up to 10 percentage points over the state-of-the-art;
\item We investigate how to correctly extract motif features and apply them to the link prediction problem;
\item We re-examine the common practice of removing existing edges from the graph to create the classification dataset, and propose an alternative based on adding negative examples which provides better accuracy;
\item We detail the effect of the distance of the pair of nodes for negative examples on the classification accuracy.
\end{squishlist}

\section{Problem Definition and Preliminaries}
\label{sec:problem}

Consider graph $G(V,E_{t_!})$ at a given time $t_1$, where $V$ is the set of nodes in the graph and $E_{t_1}$ is the set of edges that connect the nodes of the graph at that time.
Link prediction aims to predict which new edges are likely to appear at time $t_2 > t_1$,
i.e., to predict the set $\{e : e \not\in E_{t_1} \wedge e \in E_{t_2} \}$.
Clearly, the assumption is that the set of nodes $V$ does not change in time.
In addition, we assume the graph $G$ is undirected and unweighted.

While the real application of link prediction involves time, very often testing prediction algorithms in these conditions is not straightforward, mostly due to the unavailability of the history of the evolution of the graph structure.
Therefore, in most cases, link prediction is cast as a standard binary supervised classification task~\cite{al2006link}.
In this scenario, each data point corresponds to a pair of nodes $(u,v)$ in the graph, and the label 
\[
    L(u,v) = \begin{cases}
	+1	& (u,v) \in E , \\
	-1	& (u,v) \not\in E .\\
    \end{cases}
\]

To build a dataset for classification, we need a training set \trainset and a test set \testset.
Usually, a small fraction of the edges in the graph $\testset^{+} \subset E$ are kept aside for testing,
and the prediction model is trained on (part of) the rest of the edges in the graph $\trainset^{+} \subseteq E \setminus \testset^{+}$.
Often, to obtain a better statistical evaluation of the predictor, this procedure is repeated via cross-validation.

While the edges in the graph can be used for the positive examples, a binary classifier also needs negative examples.
We can extract a set of negative examples by sampling pairs of nodes from the graph which are not connected by an edge, i.e., $\trainset^{-} \subset \{(u,v) : u,v \in V \wedge(u,v) \not\in E\}$, and similarly for the test set $\testset^{-}$.
We call these pairs \emph{negative edges}.

The major challenge of this approach is to choose suitable features for the classification task.
We address this point in the next section, where we describe the motif features we propose.
Our hypothesis is that more complex topological features, which look deeper in the structure of the neighborhood of two nodes,
are more powerful predictors of link formation than features that rely on
shallow structural information such as common neighbors.
However, an equally important problem is how to create the classification dataset,
and in particular, how to sample the negative edges.
This problem has not received much attention in the literature so far, but we show that it has a significant impact on classification accuracy.

\subsection{Motifs}
\label{sec:Motifs}

\begin{figure}[b]
	\centering
	\includegraphics[width=\columnwidth]{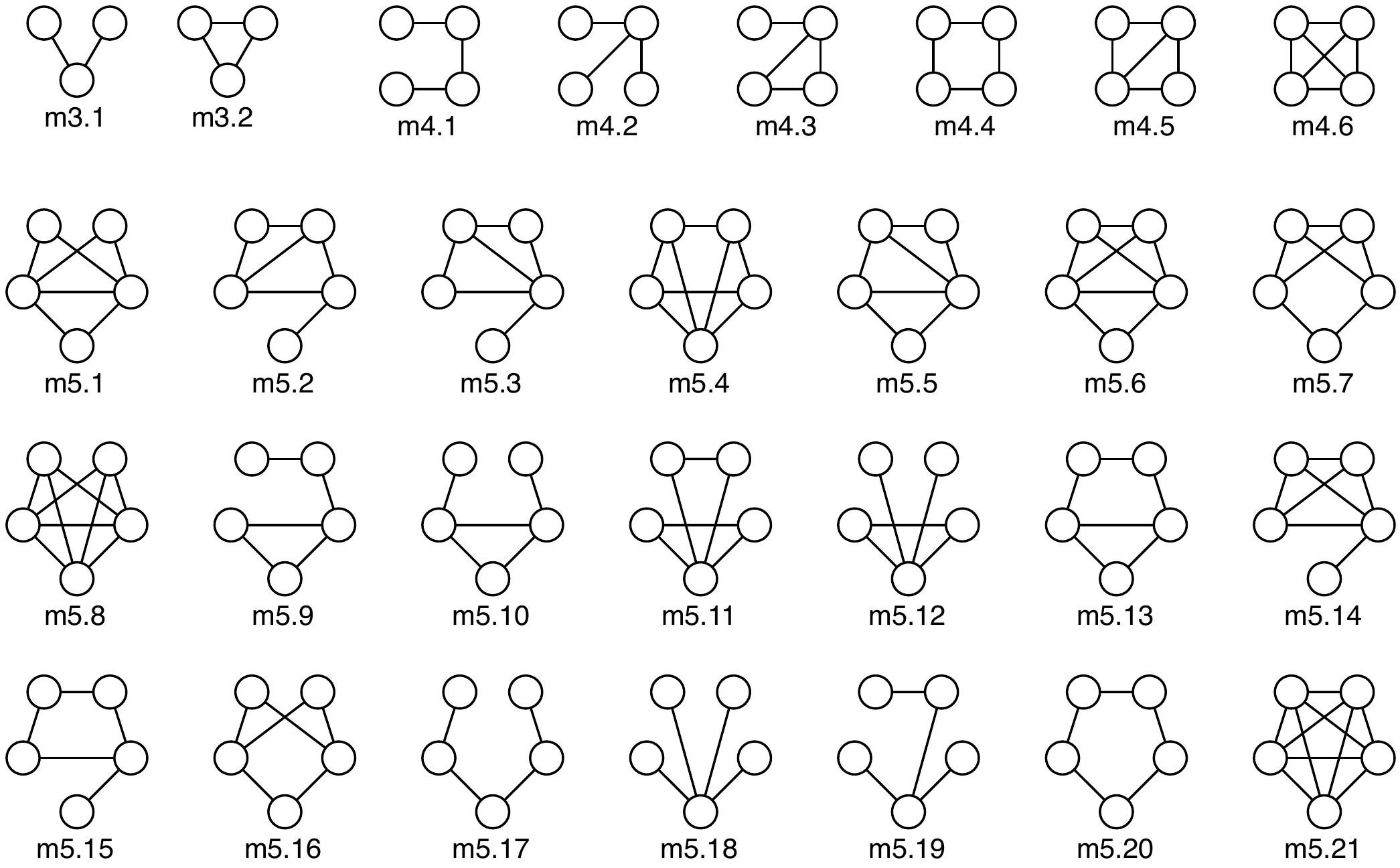}
	\caption{All the possible undirected $k$-motifs for $\mathbf{k \in \{3,4,5\}}$. There are 2 different 3-motifs, 6 different 4-motifs, and 21 different 5-motifs.}
	\label{fig:motifs}
\end{figure}

Motifs are small, connected, non-isomorphic subgraphs which appear in a larger graph~\cite{milo2002network,shen2002network}.
Each $k$-motif represents a topological pattern of interconnection between $k$ nodes in a graph.
Figure~\ref{fig:motifs} shows all the $29$ possible motifs for $k \in \{3,4,5\}$.
We denote each motif as `m$k$.$n$' where $k$ is the number of nodes in the motif and $n$ is an ordinal number which identifies the specific edge pattern in the motif
(we use this naming of motifs throughout the paper).

Motifs differ from the related concept of graphlets since they are not induced (i.e., they can be partial subgraphs), while graphlets are induced subgraphs (i.e., there is a bijection between the graphlet and its occurrence in a graph)~\cite{prvzulj2004modeling}.
Motifs are often associated with a specific frequency threshold, which needs to be higher than a null-model randomized graph.
We do not impose such restriction in the current work, and let the machine learning algorithm determine whether a particular motif is significant for our task.
The concept of motifs was originally introduced by~\cite{milo2002network}, which showed their modeling power in biological networks.

Motifs have been shown to be a powerful graph analysis tool in previous work.
The motif profile, the frequency distribution of the motifs in a graph, is used as a `fingerprint' of a graph~\cite{milo2004superfamilies}.
Therefore, the usefulness of motifs to capture the macro structure of a graph is well established~\cite{vazquez2004topological}.
However, for our purpose, we are more interested in their ability to capture the micro and meso structure (i.e., the neighborhood) of the graph.
Given that features based on triangle closing (i.e., common neighbors) are already quite effective, we expect features based on $k$-motifs for $k>3$ to hold even more predictive power.

Counting $k$-motifs is an expensive operation, as their number grows exponentially in $k$.
However, thanks to recent advances in both algorithms and systems, we are now able to count $k$-motifs on graphs with millions of edges for values of $k$ of $5$ or more~\cite{bressan2017counting,teixeira2015arabesque}.
We leverage this capability to capture complex topological features for the link prediction task, and go beyond the simple triangle-based features that have been traditionally used.


\section{Motif Features}
\label{sec:method}

The features in our model correspond to the number of occurrences of an edge (positive or negative) within different $k$-motifs.
That is, for each example edge 
in the classification dataset,
we enumerate the $k$-motifs that the edge is part of, and then count the occurrences of each different motif.
In this paper, we use 3-, 4-, and 5-motifs.
Motifs of even higher order are prohibitively expensive to compute for large graphs,
and we experimentally demonstrate high prediction accuracy with $k \in \{3,4,5\}$.
There are $2$, $6$, and $21$ motifs for $k = 3$, $4$, and $5$, respectively,
and this is the number of features we generate for each $k$.

\subsection{Equal Treatment of Positive and Negative Examples}

It is of paramount importance to treat both positive and negative example edges
in the same way with respect to feature extraction, especially when dealing with the test set.
To exemplify why this is important, imagine using $k=3$ and not addressing this issue.
The two possible features are then the wedge (or open triangle) and the closed triangle.
Positive edges
will have a mix of both features, but
negative edges
will never appear in a closed triangle, by construction.
Thus, this way of extracting features leaks information about the class into the features themselves.
This leakage is clearly an issue for the test set, but in order for the features to be meaningful, we need to apply the same extraction process to both the training set and the test set.

\begin{figure*}[t]
   \centering
\begin{tabular}{ccccccccc}
	\toprule
	& & &
	\multicolumn{6}{c}{\textbf{Features}}  \\
	\cmidrule{4-9}
	& & & 
	\includegraphics[width=2em]{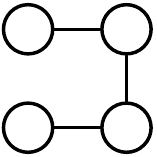} \quad &
	\includegraphics[width=2em]{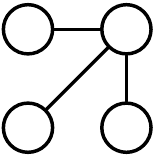} \quad &
	\includegraphics[width=2em]{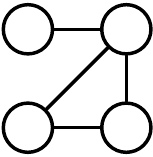} \quad &
	\includegraphics[width=2em]{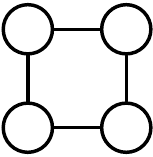} \quad &
	\includegraphics[width=2em]{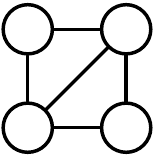} \quad &
	\includegraphics[width=2em]{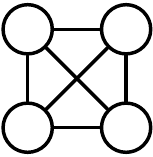} \\ 
	\midrule
	Graph &  Edge  &   Class   & m4.1 & m4.2 & m4.3  & m4.4 & m4.5 & m4.6  \\
	\midrule
	\multirow{3}{*}{\includegraphics[width=3em]{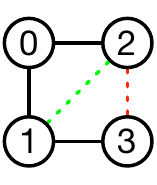}} \\[-4pt] 
	& 1-2  &  Positive & 1 & 0 & 0 & 0 &  0 & 0  \\[3pt]
	& 2-3  &  Negative  & 1  & 0 & 0 & 0 & 0 & 0  \\[3pt]
	\bottomrule
\end{tabular}
    \caption{Motif features when positive examples are removed from the graph (\rmve).}
    \label{fig:remove}
    \vspace{-0.5\baselineskip}
\end{figure*}

\begin{figure*}[t]
   \centering
\begin{tabular}{ccccccccc}
	\toprule
	& & &
	\multicolumn{6}{c}{\textbf{Features}}  \\
	\cmidrule{4-9}
	& & &
	\includegraphics[width=2em]{images/m41.pdf} \quad &
	\includegraphics[width=2em]{images/m42.pdf} \quad &
	\includegraphics[width=2em]{images/m43.pdf} \quad &
	\includegraphics[width=2em]{images/m44.pdf} \quad &
	\includegraphics[width=2em]{images/m45.pdf} \quad &
	\includegraphics[width=2em]{images/m46.pdf} \\
	\midrule
	Graph &  Edge  &   Class   & m4.1 & m4.2 & m4.3  & m4.4 & m4.5 & m4.6   \\ \midrule 
	\multirow{3}{*}{\includegraphics[width=3em]{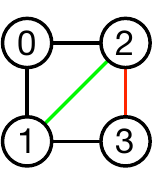}} & & & &  &  & & \\[-4pt] 
	& 1-2  &  Positive & 2 & 2 & 4 & 0 & 1 & 0  \\[3pt]
	& 2-3  &    Negative & 4 & 1 & 3 & 1 & 1 & 0 \\[3pt]
	\bottomrule
    \end{tabular}
    \caption{Motif features when negative examples are inserted into the graph (\inse).}
    \label{fig:insert}
    \vspace{-0.5\baselineskip}
\end{figure*}

To solve this problem
we have two possible options:
($i$)~remove positive edges from the motif, which we denote \rmve, 
or ($ii$)~insert negative edges into the motif, which we denote \inse.
The former option corresponds to the traditional way of handling link prediction as a classification task, where a set of (positive) edges are withheld from the model.
The latter is a 
novel way of handling the feature extraction that has not been considered previously.
It corresponds to asking the following question: ``If this edge was added to the graph, would its neighborhood look like other ones already in the graph?''

In the first method, \rmve, we remove the example positive edges from the graph and extract the features by looking at motifs that contain both endpoints of a 
removed edge.
The features for negative edges are computed in a similar manner, by looking at the motifs containing both endpoints of the negative pair.
In this case, no modification to the graph is needed for negative edges.
By following this methodology, a number of motifs will never appear 
as features (e.g., fully connected cliques).
In addition, an example edge 
never contributes to producing the motifs that it is part of.
An example for $k=4$ is shown in Figure~\ref{fig:remove}.
Let green edges be positive examples, red edges be negative examples,
and black edges be part of the graph but not in the
classification dataset (i.e., not sampled).
Additionally, dashed edges are removed from the graph.
In this case, positive edge $(1,2)$ is sampled for the classification dataset and removed from the graph,
and negative edge 
$(2,3)$ is sampled but not inserted.
Removing edge $(1,2)$ changes the motifs in this neighborhood.
For example, motifs m4.2 and m4.3 do not appear 
even though edge $(1,2)$ was part of instances of these motifs in the original unmodified graph.
After removing edge $(1,2)$, the only 4-motif that appears is m4.1, which appears once.
Since it contains the nodes in edges $(1,2)$ and $(2,3)$,
both edges have a value 1 for feature m4.1.

In the second method, \inse, we
insert negative example edges into the graph before extracting and counting motifs.
No modification to the graph is needed for positive example edges.
After inserting the negative example edges, we count the motifs for positive and negative edges in the same way.
%
All motifs can appear as features, and an example edge contributes to all the motifs it is part of.
Figure~\ref{fig:insert} shows the same example as Figure~\ref{fig:remove},
but now the negative edge $(2,3)$ is added to the graph.
Each feature of an example edge (positive or negative) corresponds to a motif which includes the edge itself.

As an illustration of extracting motif features, consider m4.2 and m4.3 in Figure~\ref{fig:insert}.
Motif m4.2 occurs twice in the graph, $(0,1)$-$(1,2)$-$(1,3)$ and $(0,2)$-$(1,2)$-$(2,3)$.
Both occurrences contain edge $(1,2)$ while only one contains edge $(2,3)$, 
so edge $(1,2)$ has a value 2 for feature m4.2 while edge $(2,3)$ has a value 1.
There are four occurrences of motif m4.3 in the graph, obtained by removing one of the edges $(0,1)$, $(1,3)$, $(0,2)$, or $(2,3)$. All of these occurrences include edge $(1,2)$ but only three include edge $(2,3)$, 
so edge $(1,2)$ has a value 4 for feature m4.3 while edge $(2,3)$ has a value 3.

When using the \inse method, we insert all of the negative edges in the graph before doing any feature extraction.
Sampling a negative edge in the neighborhood of a positive one changes the extracted features, as shown in Figure~\ref{fig:insert}.
That is, the extracted motifs are not fully independent of the sampling.
While this is not desirable, we verify that the occurrence of these cases in practice is very rare,
so they do not affect the learning process in any significant way.

\subsection{Sampling Negative Edges}

Another important question, independent of choosing \rmve or \inse, is
how to sample the edges for the classification dataset.
For positive
example
edges, uniform random sampling is an adequate solution,
reflecting the assumption that no edge is easier or harder to predict than another.
For negative
example
edges, however, it is easy to imagine that an edge connecting two nodes in completely different regions of the graph is less likely to 
occur
than one connecting two nodes in the same 
region.
Therefore, the distance between the pair of sampled nodes can play an important role.
For this reason, we choose to control
this parameter.

\begin{figure}
\centering
\subfloat[Negative edge $(1,2)$ connects nodes at distance 2.]
{\includegraphics[width=0.35\columnwidth]{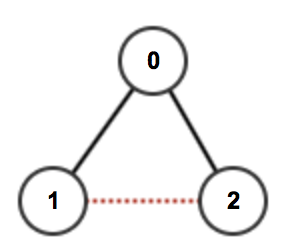}}
\hspace{1cm}
\subfloat[Negative edge $(2,3)$ connects nodes at distance 3.]
{\includegraphics[width=0.35\columnwidth]{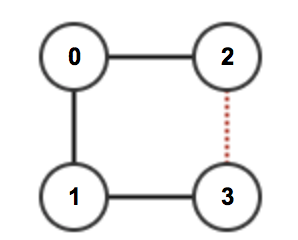} }
\caption{Two negative edges (dashed red lines) at different distances.}
\label{fig:neg-edge-dist}
\end{figure}

We sample negative edges based on the shortest-path distance between the endpoint nodes.
We choose to use a mix of nodes with distance $d=2$ and $d=3$ hops.
Figure~\ref{fig:neg-edge-dist} shows
examples of sampling negative edges with these two 
distances.
In most of the experiments, we use a 50/50 split between negative example edges at distance 2 and 3.
However, we also analyze the effect of the distance on classification accuracy by changing the ratio between these two sub-classes.

When building the classification dataset, we sample an equal number of negative and positive edges.
This decision allows us to use simple classification measures, such as accuracy, without the issues that arise due to class skew.
In a typical graph, most pairs of nodes would not have an edge connecting them, so 
the negative class would be much larger than the positive class.
However, as we are only interested in the relative performance of the features, and because we use off-the-shelf classifiers,
we prefer to create a balanced classification dataset.


\subsection{Classification}

The discussion up to this point enables us to create a classification dataset with an equal number of
positive and negative examples.
The negative examples are pairs of nodes with a shortest-path distance of 2 or 3
(typically split 50/50 between the two distances).
Each example has a set of features based on its $k$-motifs, where $k$ is 3, 4, 5, or a combination thereof.

This classification dataset can be used as input to any machine learning classification algorithm.
We experiment with several off-the-shelf classifiers, and we show in 
the following section that 
this method of creating a classification dataset combined with a powerful classification model leads to very high prediction accuracy.

\pagebreak
\section{Experiments \& Evaluation}
\label{sec:exp}
This section describes how we apply and evaluate the motif features for
link prediction on  
three different graphs.
In particular, we look at the predictive power of motif features for different motif sizes.
We also compare the different methodologies for feature extraction described in Section~\ref{sec:method}.


\subsection{Datasets}
\label{sec:dataset}
We use three real-world graphs coming from different domains: Amazon, CondMat, and AstroPh.
The three 
graphs are from the Koblenz Network Collection.\footnote{\url{http://konect.uni-koblenz.de}}
Table~\ref{tab:dataset} shows basic statistics about these graph datasets.

The first graph represents the co-purchase network of products on Amazon.
It is the graph upon which the ``customers who bought this also bought that'' feature is built.
The nodes are products, and an edge between any two nodes shows that the two products have been frequently bought together.
In this application domain, link prediction tries to model the creation of an association between two products.
In other words, it can help identify hidden relationships between products, which can then be used in a customer-facing recommendation system.
The second dataset, CondMat, is a graph which represents a subset of authorship relations between authors and publications in the arXiv condensed matter physics section.
Nodes are authors (first set) or papers (second set).
An edge represents that an author has written a given paper.
Link prediction can identify whether an author is a likely contributor to a research paper, for instance, to identify missing records.
The third and final dataset, AstroPh, is a collaboration graph.
In particular, it contains data about scientific collaboration between authors in the arXiv astrophysics section.
Each node in the graph represents an author of a paper, and an edge between two authors represents a common publication.
Predicting future collaborations between authors based on their previous publications is a well-known application for link prediction.

\begin{table}[tb]
  \centering
  \vspace{9mm}
  \caption{Basic statistics about the graph datasets.}
  \label{tab:dataset}
  \begin{tabular}{lrrrrrr}
	\toprule
	\textbf{Graph} & \boldmath $\lvert \textbf{V} \rvert$ & \boldmath $\lvert \textbf{E} \rvert$ & \textbf{Avg. Degree} & \textbf{Diameter} \\
	\midrule
  Amazon  & \num{334863}  & \num{925872}  & \num{5.530}    & 47 \\
  CondMat & \num{22015}   & \num{58595}   & \num{3.025}    & 36 \\
  AstroPh   & \num{18771}   & \num{198050}  & \num{21.102} & 14 \\
	\bottomrule
\end{tabular}

\vspace{8mm}

    \centering
     \vspace{7mm}
	\caption{Number of positive and negative edges sampled for the classification dataset for each graph.}
	\label{tab:sample}
	\begin{center}
	\begin{tabular}{lrr}
	\toprule
	\textbf{Graph} & \textbf{\# Positive Edges} & \textbf{\# Negative Edges}  \\
	\midrule
	Amazon	& \num{20000}	& \num{20000} \\
	CondMat	& \num{2000}	& \num{2000}   \\
	AstroPh 	& \num{5000}	& \num{5000}   \\
	\bottomrule	
	\end{tabular}
	\end{center}
\end{table}

\pagebreak
\subsection{Experimental Settings}
\spara{Classification Datasets.}
For each graph, we extract a classification dataset for which we compute node pair features.
We extract a uniform sample of edges from each graph as positive examples.
For negative examples, we extract pairs of nodes from the graph which are at distance $2$ or $3$ hops.
Table~\ref{tab:sample} shows the number of examples chosen from each graph.
As already mentioned, to assess the predictive power of the motif features in a simple setting, we build balanced classification datasets.

\spara{Feature Extraction.}
We extract the motifs in which the example edges in the classification dataset occur using the Arabesque parallel graph mining framework~\cite{teixeira2015arabesque,hussein2017arabesquedemo}.
We then group by motif, count the occurrences of each motif, and finally normalize the counts to create a feature vector for each example edge that represents the motif distribution of the neighborhood of the edge.\footnote{Code available at \url{https://github.com/GhadeerAbuoda/LinkPrediction}.}

\spara{Classification Models.}
To train the classification models we use the scikit-learn Python library~\cite{scikit-learn}.
We experiment with the following models: na\"{i}ve Bayes (NB), logistic regression (LR), decision tree (DT), k-nearest neighbors (KNN), gradient boosted decision trees (GB), and random forest (RF).
All the classification performance results are computed via 10-fold cross-validation.

\spara{Baselines.}
We use two types of baselines.
The first type includes traditional topological features such as triangle closure and paths.
We compare our features against common neighbors~\cite{newman2001clustering}, Jaccard coefficient,
Adamic/Adar measure~\cite{adamic2003friends},
preferential attachment~\cite{barabasi2009scale},
rooted PageRank,
and Katz index~\cite{katz1953new}.
Of these methods, PageRank and Katz benefit from inserting negative edges in the graph,
so we use \inse with these two methods.

\begin{sloppypar}
The second type of baseline includes more complex techniques such as matrix decomposition and deep learning.
For matrix decomposition, we use the scores obtained from a non-negative matrix factorization (NMF) trained on the graph with positive edges removed (\rmve), as commonly done 
in the literature~\cite{menon2011link}.
We use the NMF algorithm available in scikit-learn, and use $100$ factors for the decomposition.
For deep learning, we compare against a recent state-of-the-art graph neural network framework for link prediction called SEAL~\citep{zhang2018link}.
SEAL uses subgraph extraction around the example edge to extract latent features, learned via a neural network.
This framework has experimentally outperformed other existing deep learning methods such as node2vec and LINE~\citep{grover2016node2vec,tang2015line}.
\end{sloppypar}

\subsection{Evaluation Metrics}
We evaluate the classification task via the following three metrics:
\begin{squishlist}
\item Accuracy (ACC): the fraction of examples correctly classified, both true positives (TP) and true negatives (TN), over the total number of examples (N).
\[
ACC = \frac{TP+TN}{N}
\]
Given that the classification datasets are balanced, accuracy is a reasonable measure of performance.
Better classifiers obtain higher accuracy.

\item Area Under the Curve (AUC): the area under the Receiver Operating Characteristic (ROC) curve generated from the scores produced by the classifiers.
This measure represents the probability that a classifier will rank a randomly chosen positive example higher than a randomly chosen negative one. Better classifiers obtain higher AUC.

\item False Positive Rate (FPR): the ratio between the number of negative edges wrongly classified (false positives) and the total number of negative edges in the dataset.
\[
FPR = \frac{FP}{FP+TN}
\]
This measure is useful to understand the effect of the graph distance of the negative examples on the classification task.
Better classifiers obtain lower FPR.

\end{squishlist}

\subsection{Removing Positive Edges vs. Inserting Negative Edges}
Tables~\ref{tab:ACC-amz-rmv} and~\ref{tab:ACC-amz-ins} show the classification results of the two feature extraction methods (\rmve and \inse, respectively) on the Amazon dataset (the largest one).
We report the results for all classifiers when using features based only on motifs of size $k=3$, only on motifs of size $k=4$, and only on motifs of size $k=5$. We also report in the last column the results when using all three sets of features together in one feature vector (total of 29 features).

By looking at the difference between the two tables in the first six rows (showing accuracy), it is clear that \inse consistently has higher accuracy than \rmve.
The difference grows smaller as we add more complex features by increasing $k$.
However, for the two best classifiers (GB and RF), \inse
still results in approximately 3 percentage points higher accuracy than \rmve even when using the combined features.
The simpler classifiers do not seem able to exploit the full predictive power of the motif features. On the other hand, GB and RF are advanced classifiers that can exploit this predictive power.

Tables~\ref{tab:ACC-amz-rmv} and~\ref{tab:ACC-amz-ins}
report AUC and FPR
for the two best classifiers.
As with the accuracy metric, AUC and FPR show that \inse is better than \rmve.
The AUC and FPR values are very similar for GB and RF, and the two classifiers are almost indistinguishable.
As expected, the more complex motif features (i.e., larger $k$) work better, and the combination of all three sets of features is usually the best.
For ease of presentation, henceforth we report results only 
using the RF classifier, but results using GB are similar.

\begin{table}[t]
  \caption{Classification performance on Amazon (\rmve).}
  \centering
  \small
  \label{tab:ACC-amz-rmv}
  \begin{tabular}{llcccc}
\toprule
     \multirow{2}{*}{Metric} & \multirow{2}{*}{Classifier}& \multicolumn{4}{c}{Features} \\
	\cmidrule(lr){3-6}
      &  & $k=3$ & $k=4$  &  $k=5$ & Combined  \\ 
\midrule
\multirow{6}{*}{ACC (\%)} 
	& NB		& \num{57.6} & \num{52.4} & \num{52.7} & \num{52.0} \\
	& LR		& \num{56.5} & \num{59.4} & \num{68.0} & \num{64.4} \\
	& DT		& \num{57.6} & \num{69.4} & \num{70.8} & \num{70.6} \\
	& KNN	& \num{51.5} & \num{69.9} & \num{71.4} & \num{71.0} \\
	& GB	& \num{58.0} & \num{73.3} & \num{76.6} & \num{76.9} \\
	& RF		& \num{57.6} & \num{71.6} & \num{76.3} & \num{77.0} \\
\midrule
\multirow{2}{*}{AUC}
	& GB	& \num{0.58} & \num{0.72} & \num{0.76} & \num{0.76} \\
	& RF		& \num{0.58} & \num{0.72} & \num{0.76} & \num{0.77} \\
\midrule	
\multirow{2}{*}{FPR}
	& GB	& \num{0.11} & \num{0.30} & \num{0.26} & \num{0.27} \\
	& RF		& \num{0.11} & \num{0.32} & \num{0.27} & \num{0.27} \\
\bottomrule

\end{tabular}
\end{table}

\begin{table}[t]
  \caption{Classification performance on Amazon (\inse).}
  \centering
  \small
  \label{tab:ACC-amz-ins}
  \begin{tabular}{llcccc}
\toprule
     \multirow{2}{*}{Metric} & \multirow{2}{*}{Classifier} & \multicolumn{4}{c}{Features} \\
	\cmidrule(lr){3-6}
      &  & $k=3$ & $k=4$  &  $k=5$ & Combined  \\ 
\midrule
\multirow{6}{*}{ACC (\%)}
	& NB		& \num{57.7} & \num{57.2} & \num{52.7} & \num{53.6} \\  
	& LR		& \num{62.5} & \num{67.7} & \num{70.0} & \num{67.5} \\    
	& DT		& \num{67.9} & \num{66.9} & \num{69.6} & \num{71.0} \\ 
	& KNN	& \num{63.6} & \num{66.0} & \num{64.0} & \num{65.0} \\  
	& GB	& \num{68.2} & \num{75.0} & \num{76.6} & \num{79.4} \\
	& RF		& \num{68.0} & \num{74.8} & \num{77.0} & \num{79.6} \\ 
\midrule
\multirow{2}{*}{AUC}
	& GB	& \num{0.69} & \num{0.74} & \num{0.76} & \num{0.80} \\ 
	& RF		& \num{0.68} & \num{0.75} & \num{0.78} & \num{0.80} \\
\midrule
\multirow{2}{*}{FPR}
	& GB	& \num{0.25} & \num{0.25} & \num{0.25} & \num{0.18} \\ 
	& RF		& \num{0.23} & \num{0.23} & \num{0.21} & \num{0.18} \\ 
\bottomrule
\end{tabular}
\end{table}

\begin{figure}[t]
\centering
\subfloat[Amazon]
{\includegraphics[width=\columnwidth]{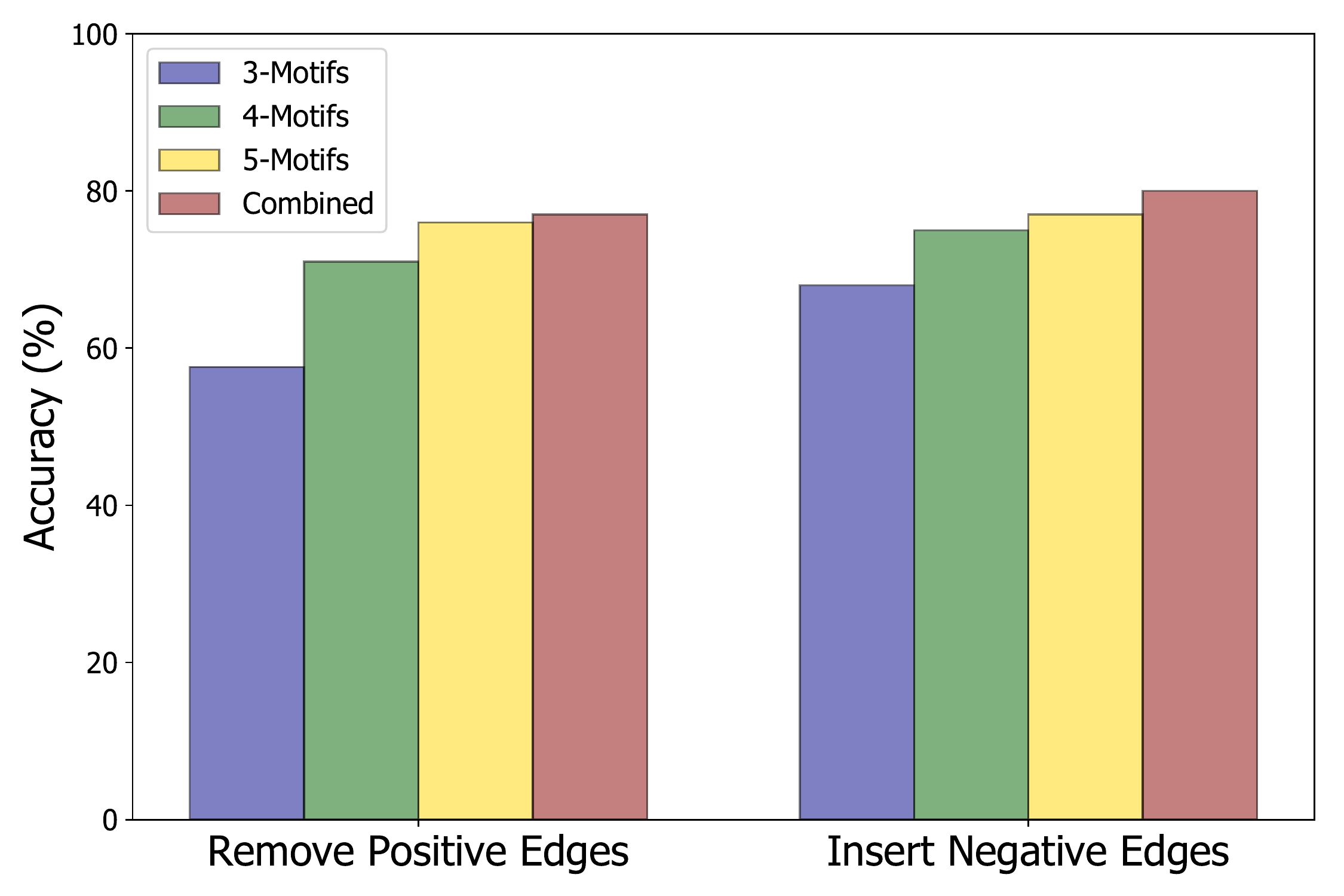}}  \\
\subfloat[CondMat]
{\includegraphics[width=\columnwidth]{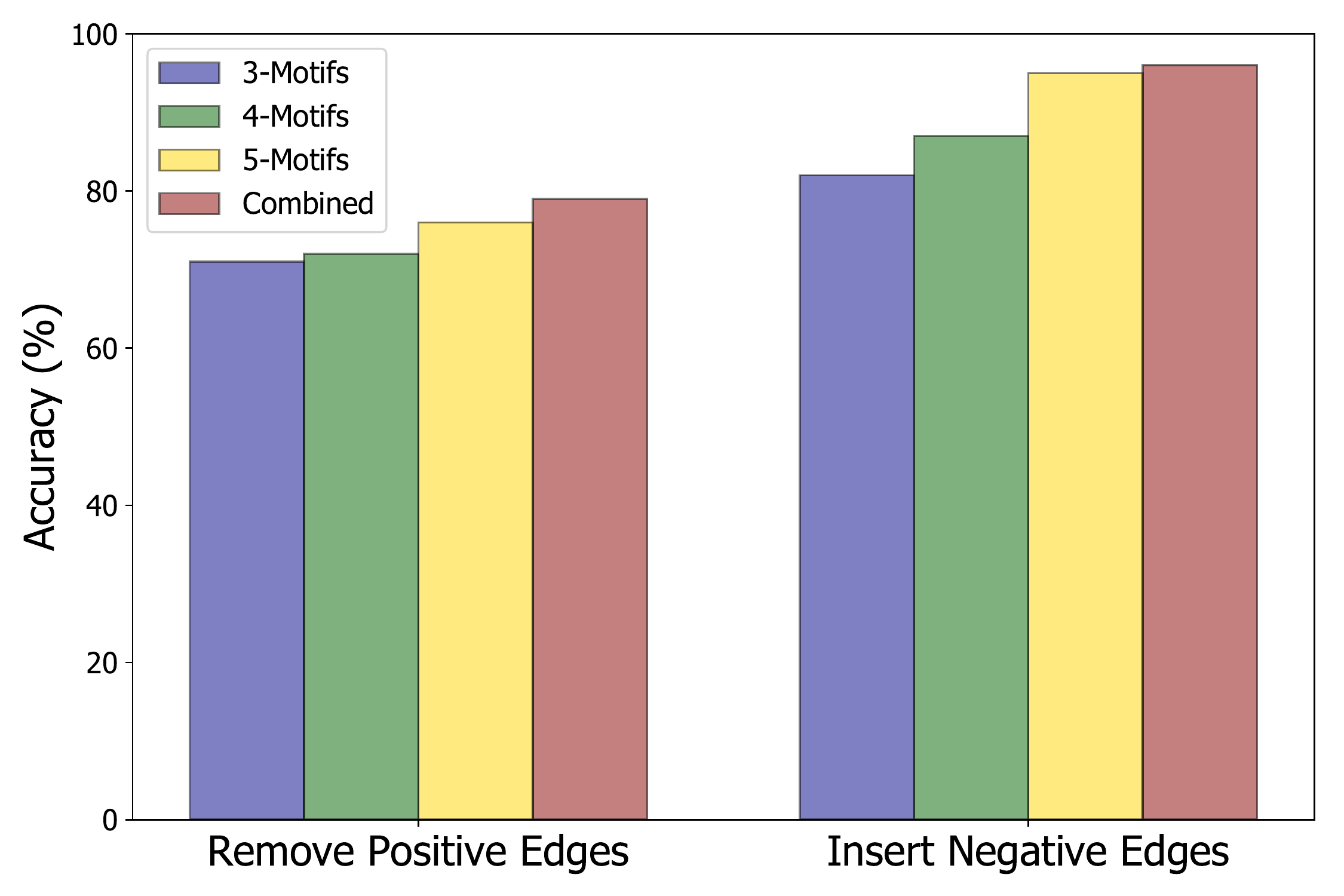}} \\
\subfloat[AstroPh]
{\includegraphics[width=\columnwidth]{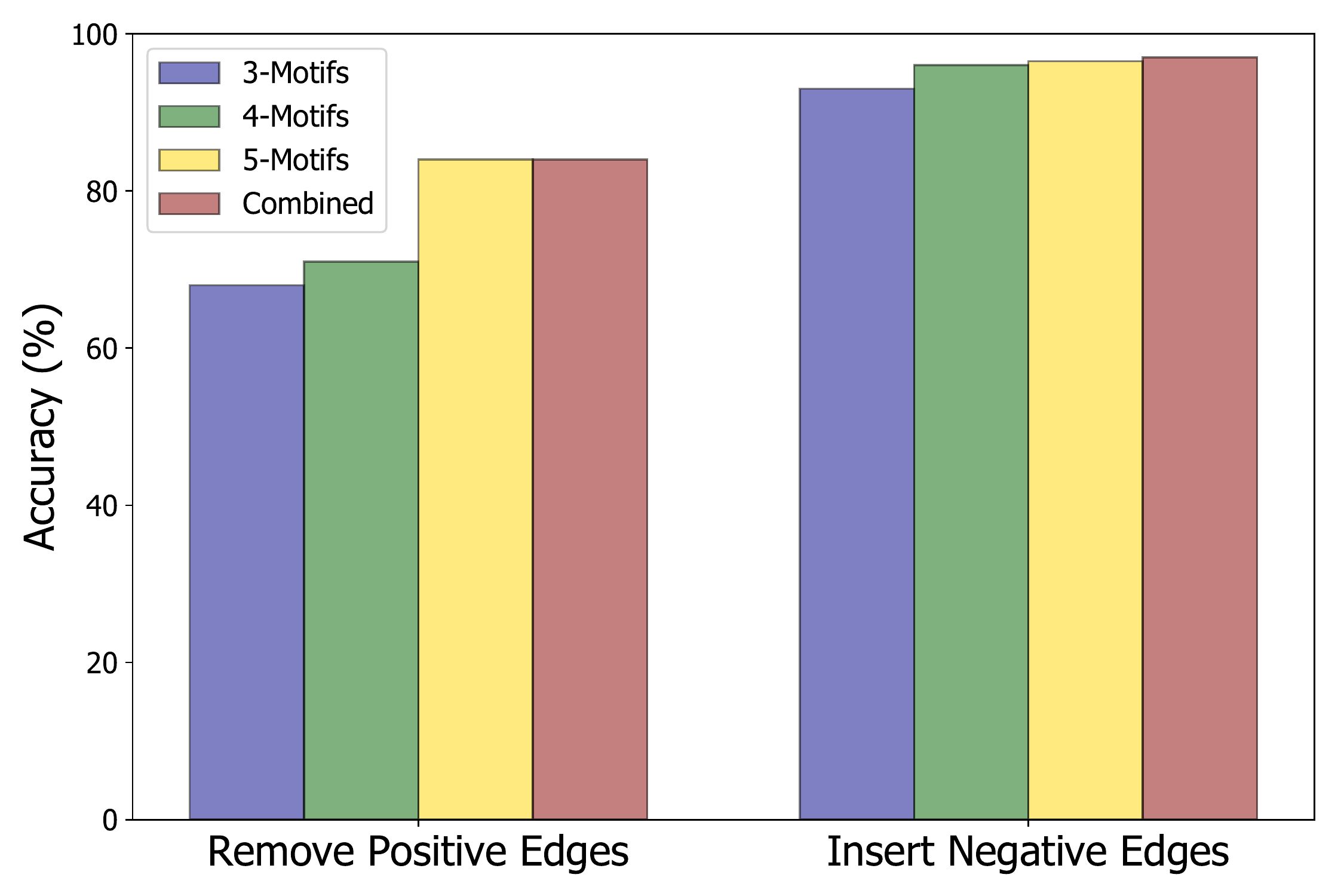}} \\
\caption{Classification accuracy of a Random Forest (RF) classifier when using different motif features and different feature extraction methods (\rmve vs.\ \inse).} 
\label{fig:accuracy} 
\end{figure} 

Figure~\ref{fig:accuracy} reports the accuracy numbers for RF on all three datasets for both feature extraction methods.
The results are consistent with what we observed above: \inse is consistently better than \rmve.
Interestingly, \inse with just 3-motif features performs better than \rmve with the combined motif features on the CondMat and AstroPh datasets.

We perform a statistical test to compare the classification accuracy of the two methods, \inse and  \rmve.
We obtain $100$ different samples of the accuracy for each method by training the RF classifier using different seeds for the pseudo-random number generator.
We use Student's t-test to compare the results, and we are able to reject the null hypothesis that the two methods have the same average performance at the $p = 0.05$ significance level.
It is clear from all the previous results that the \inse feature extraction method is superior, and we return to the reason behind this in Section~\ref{sec:distribution}.

Regarding accuracy with varying $k$, the results show that more complex motif features perform better,
with the combination of all motif sizes outperforming each individual size.
The latter result might seem surprising, as one would expect the 5-motif features to 
subsume
the smaller ones.
However,
consider that our 5-motif features do not encode positional information, i.e., we do not know in which part of the 5-motif the edge appears.
Smaller features can provide this information, thereby supplementing the 5-motif features.
For example, if a sample edge presents feature m5.3 from Figure~\ref{fig:motifs}, it could be any of that motif's edges.
But if the edge also presents no instances of feature m3.2, then we can pinpoint it to be the only the edge in the m5.3 motif that is not involved in any triangle, i.e., the bottom right edge.

Finally, Table~\ref{tab:class-perf} reports ACC, AUC, and FPR for RF on all datasets for the two different feature extraction methods when using the combined motif features.
The mix of RF, combined motif features, and \inse feature extraction is the one that performs consistently on top.
Therefore, we use it when comparing our proposal with baseline methods in the following section.

\begin{table}[tbp]
  \caption{Classification performance of a Random Forest (RF) classifier with combined motif features when using different feature extraction methods (\rmve vs.\ \inse).}
  \label{tab:class-perf}
  \centering
  \begin{tabular}{llrrr}
\toprule
    \textbf{Method} & \textbf{Graph} & \textbf{ACC}  & \textbf{AUC} & \textbf {FPR} \\
\midrule
\multirow{3}{*}{\rmve}
	& Amazon		& \num{0.770} & \num{0.77} & \num{0.27} \\
	& CondMat	& \num{0.790} & \num{0.79} & \num{0.04} \\
	& AstroPh		& \num{0.840} & \num{0.84} & \num{0.30} \\ 
\midrule
\multirow{3}{*}{\inse} 
	& Amazon		& \num{0.796} & \num{0.80} & \num{0.18} \\
	& CondMat	& \num{0.960} & \num{0.96} & \num{0.04} \\
	& AstroPh		& \num{0.965} & \num{0.97} & \num{0.02} \\
\bottomrule
\end{tabular}
\end{table}

\begin{table*}[t]
  \centering
  \caption{Accuracy of the RF classifier (\%) when using combined motif features (with \inse) vs.\ baseline classifiers.}
  \label{tab:comparison}
  \begin{tabular}{l r r r}
   \toprule
   \textbf{Features} &	\textbf{Amazon} &	\textbf{CondMat} & \textbf{AstroPh} \\
   \midrule
   Common Neighbors             & 64.6  & 78.6  & 81.2 \\
   Jaccard Coefficient								& 61.7 	& 81.1	& 85.2 \\
   Adamic/Adar	 								& 61.5	& 74.7	& 75.0 \\
   Preferential Attachment  						& 55.0   	& 61.2   	& 64.2 \\
   Rooted PageRank 							& 53.2	& 62.0	& 65.0 \\ 
   Katz Index			 	                             		& 60.0	& 55.0	& 59.0 \\
   Topological Combined  							& 73.0	& 86.9	& 87.0 \\
   \midrule
   NMF										& 52.0	& 54.0	& 53.5 \\
   NMF + Topological Combined						& 73.0	& 85.9	& 89.0 \\
   SEAL										& 69.0 	& 81.3	& 80.3 \\
   SEAL + node2vec Embeddings					& 62.8	& 77.2	& 82.0 \\
   \midrule
   Motif Combined (\inse)							& \textbf{79.6}	& \textbf{96.0}	& \textbf{96.5} \\
   \bottomrule
   \end{tabular}
\end{table*}

\subsection{Comparison with Baselines}
To compare against the baseline topological features proposed in prior work, we train a RF on each of these features, and one on the combination of all of the features.
The upper part of Table~\ref{tab:comparison} reports the accuracy of these classifiers. 
For comparison, the accuracy of the RF classifier trained on the combined motif features extracted via \inse is reported in the last row of the table.
The first four rows of the table show simple neighborhood-based topological features. 
The next two rows show path-dependent topological features. 
For rooted PageRank, we use the standard value for the damping parameter $\alpha = 0.85$.
For the Katz index, we optimize the value of the $\beta$ parameter and we report the highest accuracy obtained (which was at $\beta = 0.1$).
The accuracy of the topological features is in the range $55$--$85\%$. 
Combining all topological features into one feature vector results in the best accuracy in all cases.
This is expected since each of these features captures different information about the graph and a powerful classifier such as RF is able to exploit all of this information.
Thus, the Topological Combined row in Table~\ref{tab:comparison} can be viewed as the best possible accuracy with current state-of-the-art topological features.
We observe that
the motif features achieve much higher accuracy (last row of the table).
Specifically, they are 
$7$ to $10$ percentage points better in accuracy, which is significant given that advanced features extracted via graph embeddings and deep learning reportedly struggle to beat the traditional topological features~\cite{tsitsulin2018verse}.

Next, we turn our attention to feature learning methods that allow the model to determine by itself which features are important for link prediction. As mentioned earlier, we focus on two popular approaches: non-negative matrix factorization (NMF) and deep learning.
Interestingly, the NMF approach~\cite{menon2011link} is not very competitive, as shown in Table~\ref{tab:comparison}.
We hypothesize that the method requires more parameter optimization (e.g., tuning the number of factors used and the regularization parameters).
In any case, the gap between NMF and straightforward topological features is quite large, which is quite disappointing.
Moreover, adding the NMF features to the topological ones does not improve accuracy by much (only the AstroPh dataset sees some improvement).

Finally, we compare our model with SEAL~\cite{zhang2018link}, a recent link prediction framework which uses deep learning (graph neural networks).
We test the framework with its default hyperparameters. 
Interestingly, 
SEAL only achieves around $70\%$ accuracy on Amazon and $80\%$ accuracy on the other two datasets.
SEAL learns on one- or two-hop subgraphs extracted around the tested edge, which is somewhat equivalent to looking into common neighbors. 
However, the accuracy achieved by SEAL is lower than with the combined topological features.

We also test combining the subgraph features with node representations learned via node2vec~\cite{grover2016node2vec}, as suggested by the authors of SEAL.
The accuracy with the node2vec embeddings does not improve on average, and actually drops for two of the datasets.
One interpretation of these results is that the node2vec embeddings might actually introduce noise in the node representations by looking too far into the neighborhoods of the example edges (e.g., the length of the random walks may not be appropriately tuned).

Thus, the overall takeaway from Table~\ref{tab:comparison} is that RF with motif features is more accurate than all the baselines, both traditional topological-based ones and more recent NMF and deep learning ones.

\begin{figure}
\centering
\subfloat[Amazon]
{\includegraphics[width=\columnwidth]{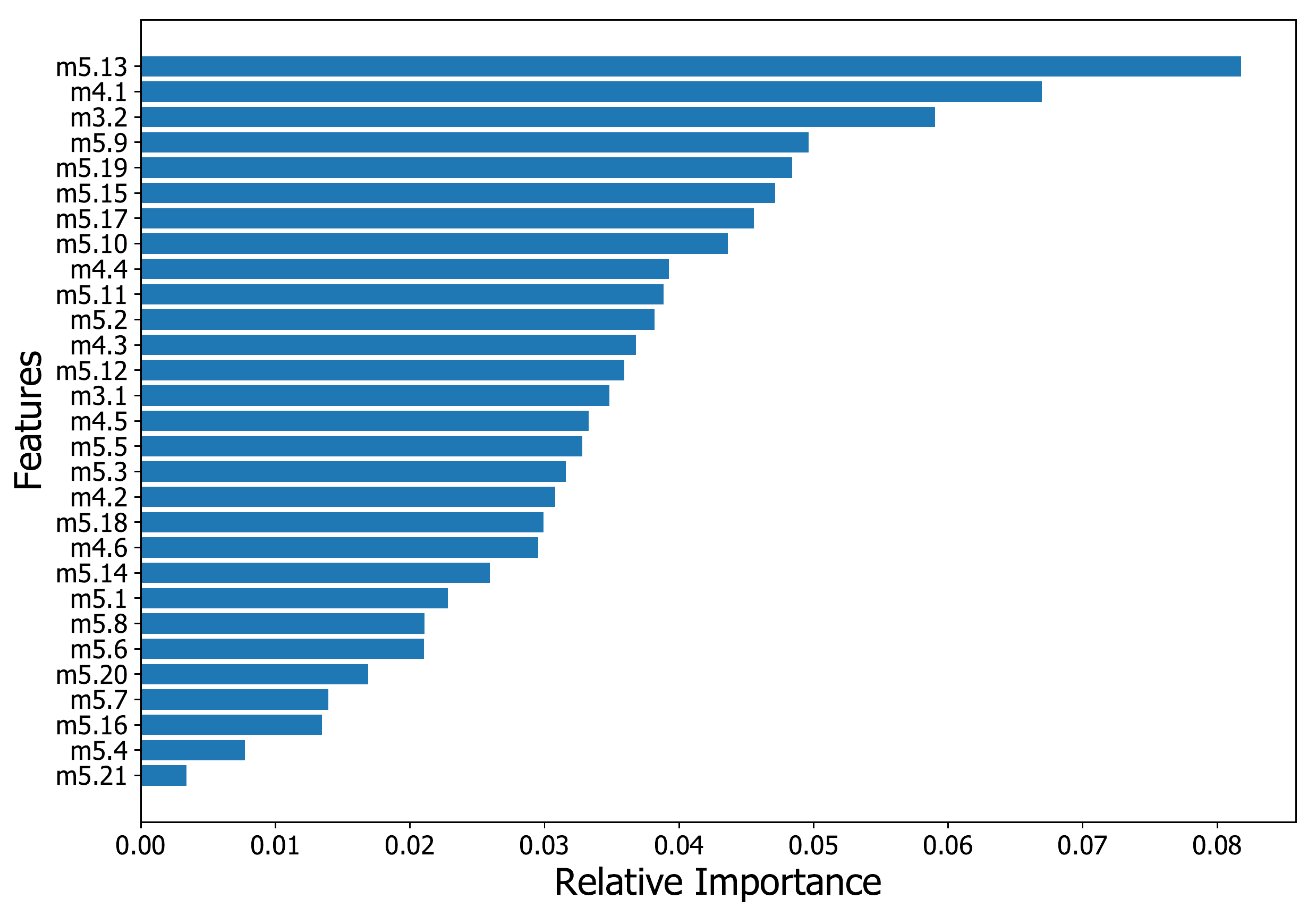}}  \\
\subfloat[CondMat]
{\includegraphics[width=\columnwidth]{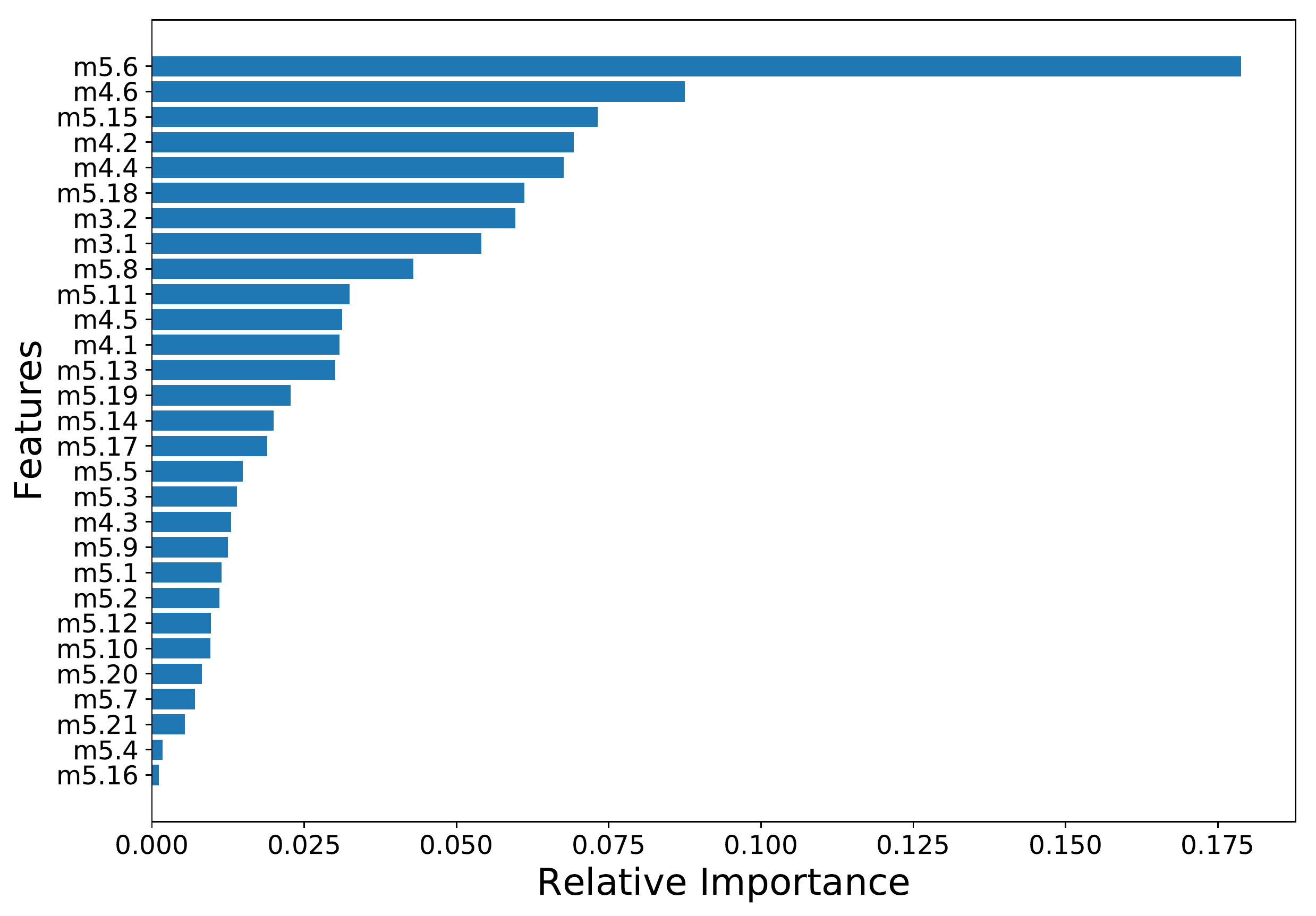}} \\
\subfloat[AstroPh] 
{\includegraphics[width=\columnwidth]{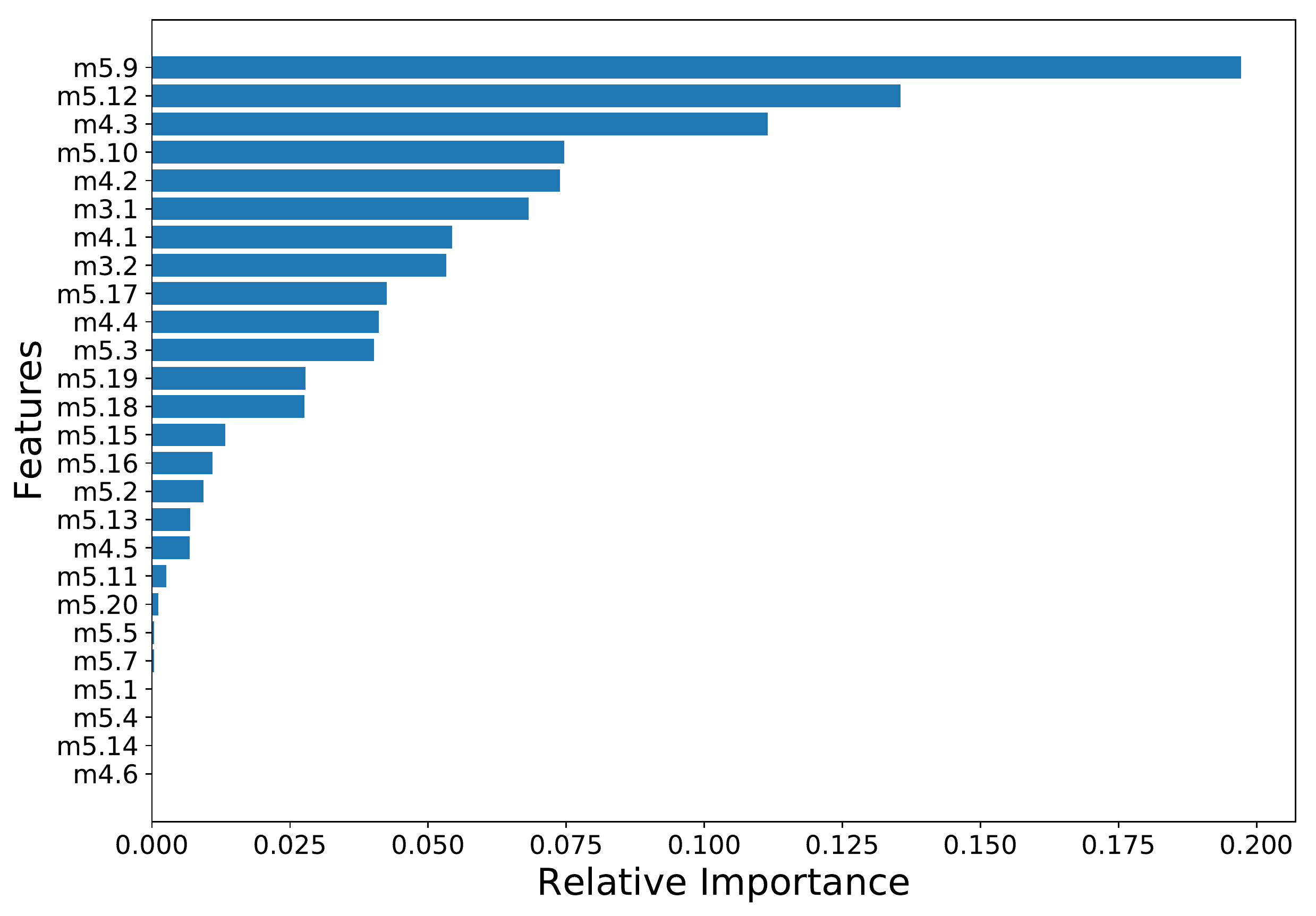}} \\
\caption{Feature importance across the three datasets as inferred from the Random Forest model. In all three datasets the most important feature is a 5-motif, but the specific motif varies by dataset.}
\label{fig:feature-importance} 
\end{figure} 

\subsection{Feature Importance}
We analyze the motif features that are most predictive for the classification task.
Figure~\ref{fig:feature-importance} shows the relative importance of the features as inferred by the Random Forest classifier. Feature importance for the RF classifier is defined by Mean Decrease in Impurity (MDI)~\cite{quinlan1987decision}. The MDI importance of a feature is calculated by measuring how effective the feature is at reducing uncertainty when creating decision trees within the RF classifier.

In all
cases the distribution of feature importance is quite skewed, with a few features constituting the backbone of the predictive model.
Interestingly, the most predictive feature is always a 5-motif one, which is another indication of the predictive power of deeper structural features.
However, the most important feature changes from dataset to dataset, and might be domain specific.

As an illustration of the utility of feature importance, we train the RF classifier with just the top-10 features by importance for each dataset,
and we observe that the accuracy obtained is almost identical to the full model.
This result indicates that feature selection could be used to identify a few important motifs in a dataset, and extract only those which would reduce the computational burden of counting all the motifs.

Overall, these results prove the predictive power of higher-order motif-based features for link formation.
The rest of the experimental section is devoted to 
three more questions related to motif feature extraction and negative edge sampling.
First, we shed some insight about why \inse performs better than \rmve.
Second, we show the importance of choosing the right negative examples, an important factor which has been mostly overlooked in the literature thus far.
Finally, while not the main focus of the current paper, we show that motif extraction is quite scalable and thus easily applicable to medium sized graphs.

\begin{table}[tbp]
  \caption{Earth Mover's Distance (EMD) and Kullback--Leibler Divergence (KLD) between the distribution of motifs in the original graph and the one obtained by each feature extraction method, \rmve and \inse. A smaller distance indicates that the given feature extraction method is more faithful to the original graph.}
  \centering
  \label{tab:distribution}
  \begin{tabular}{l cccc}
   \toprule
   \multirow{2}{*}{\textbf{Graph}} &  \multicolumn{2}{c}{\textbf{EMD }}   &  \multicolumn{2}{c}{\textbf{KLD}}  \\
   \cmidrule(lr){2-3}    \cmidrule(lr){4-5}
    &  \multicolumn{1}{c}{\textbf{\rmve}}  &  \multicolumn{1}{c}{\textbf{\inse}}  &  \multicolumn{1}{c}{\textbf{\rmve}}   &  \multicolumn{1}{c}{\textbf{\inse}}   \\    
    \midrule 
    Amazon	& \num{0.119}	& \num{0.011}	& \num{0.007}	& \num{0.001} \\
    CondMat	& \num{1.106}	& \num{0.161}	& \num{0.533}	& \num{0.012} \\
    AstroPh		& \num{0.050}	& \num{0.529}	& \num{0.001} 	& \num{0.066} \\

    \bottomrule
\end{tabular}
\end{table}

\subsection{Motif Distribution: \rmve vs. \inse}
\label{sec:distribution}
Let us now look at the reason why \inse outperforms \rmve for feature extraction.
Consider that both 
methods change the original motifs of the graph, as they alter the graph structure during feature extraction.
One hypothesis is that the method which alters the structure the least is better, as the motif patterns it learns are also the closest to the ones found in the original graph.
To test this hypothesis, we compute the motif distribution in the original graph and in the modified graphs resulting from the modifications done by \rmve and \inse (i.e., with a fraction of edges removed or added).
We compute the motif distribution for $k=3$ and $4$ for the whole graph, and compute the distance between the original distribution and the distribution obtained by \rmve and \inse.
We use two different distance functions to perform the comparison: Earth Mover's Distance (EMD)~\cite{lavin1998feature} 
and Kullback--Leibler Divergence (KLD)~\cite{kullback1951information}. 
Table~\ref{tab:distribution} reports the results.
If our hypothesis is correct, then \inse should have a smaller distance than \rmve.
This is indeed the case for two out of three graphs, for both distance functions, which gives us confidence that our hypothesis is a step in the right direction.
However, AstroPh behaves differently, with \rmve having a larger distance than \inse.
Therefore, we cannot draw a 
definitive conclusion,
and further study is necessary to fully understand the difference between these two motif feature extraction methods.

\subsection{Effect of Distance on Negative Edges}
In this experiment we explore the effect of the distance between the node pairs that constitute the negative examples on the accuracy of the classifier.
For each graph, we create different classification datasets by varying the composition of the negative class:
from containing only negative edges at distance 3 to containing only negative edges at distance 2.
We use the fraction of negative edges of the sub-class at distance 2 as the independent variable in the plots, varying it from 0 to 100\% (the rest of the edges are at distance 3).
We keep the total number of examples fixed to maintain the balance between positive and negative classes.

\begin{figure}
\centering
\includegraphics[width=\columnwidth]{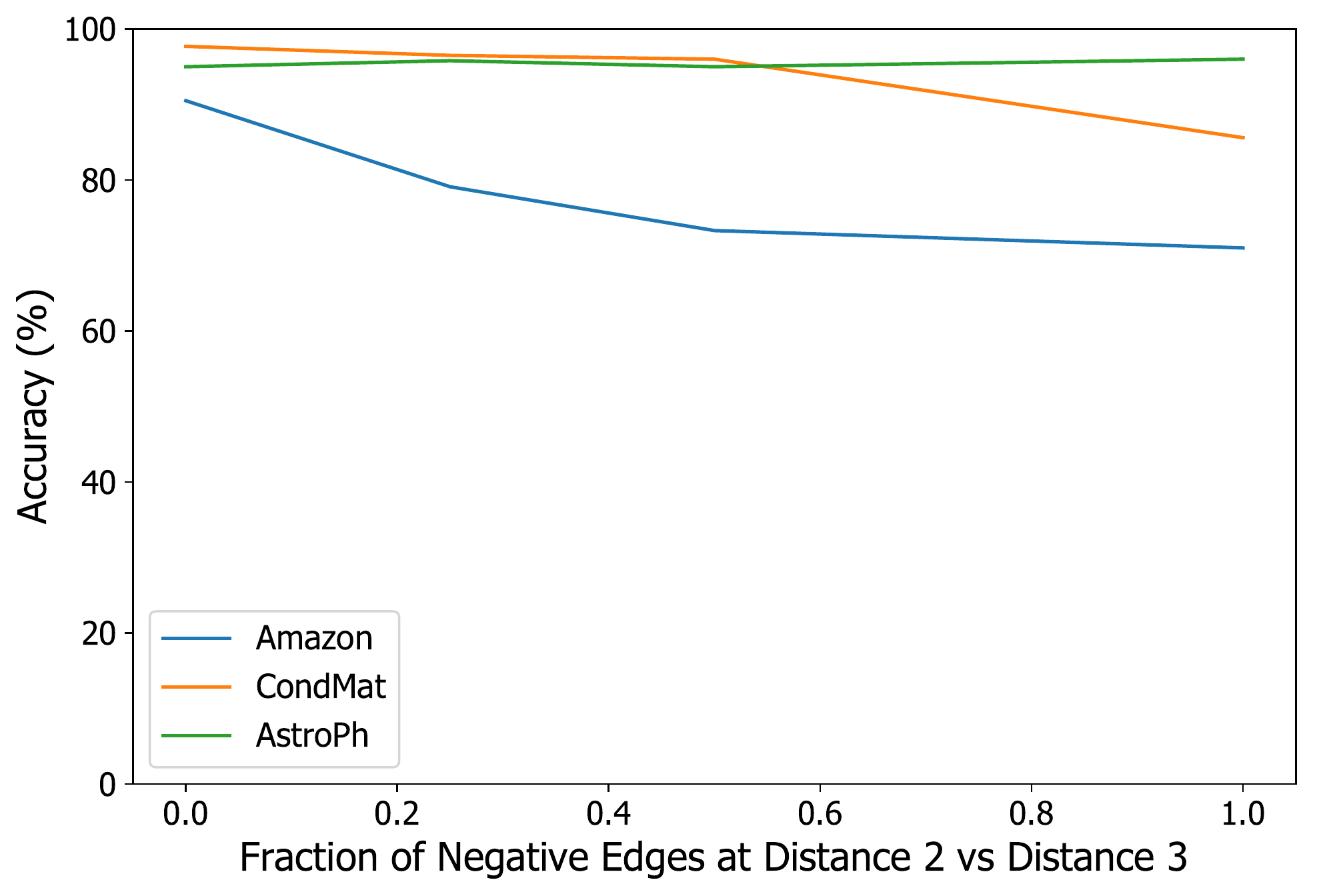}
\caption{Classification accuracy as a function of the fraction of negative examples at distance 2 (vs.\ distance 3).}
\label{fig:acc-2to3}
\end{figure}

\begin{figure}
\centering
\includegraphics[width=\columnwidth]{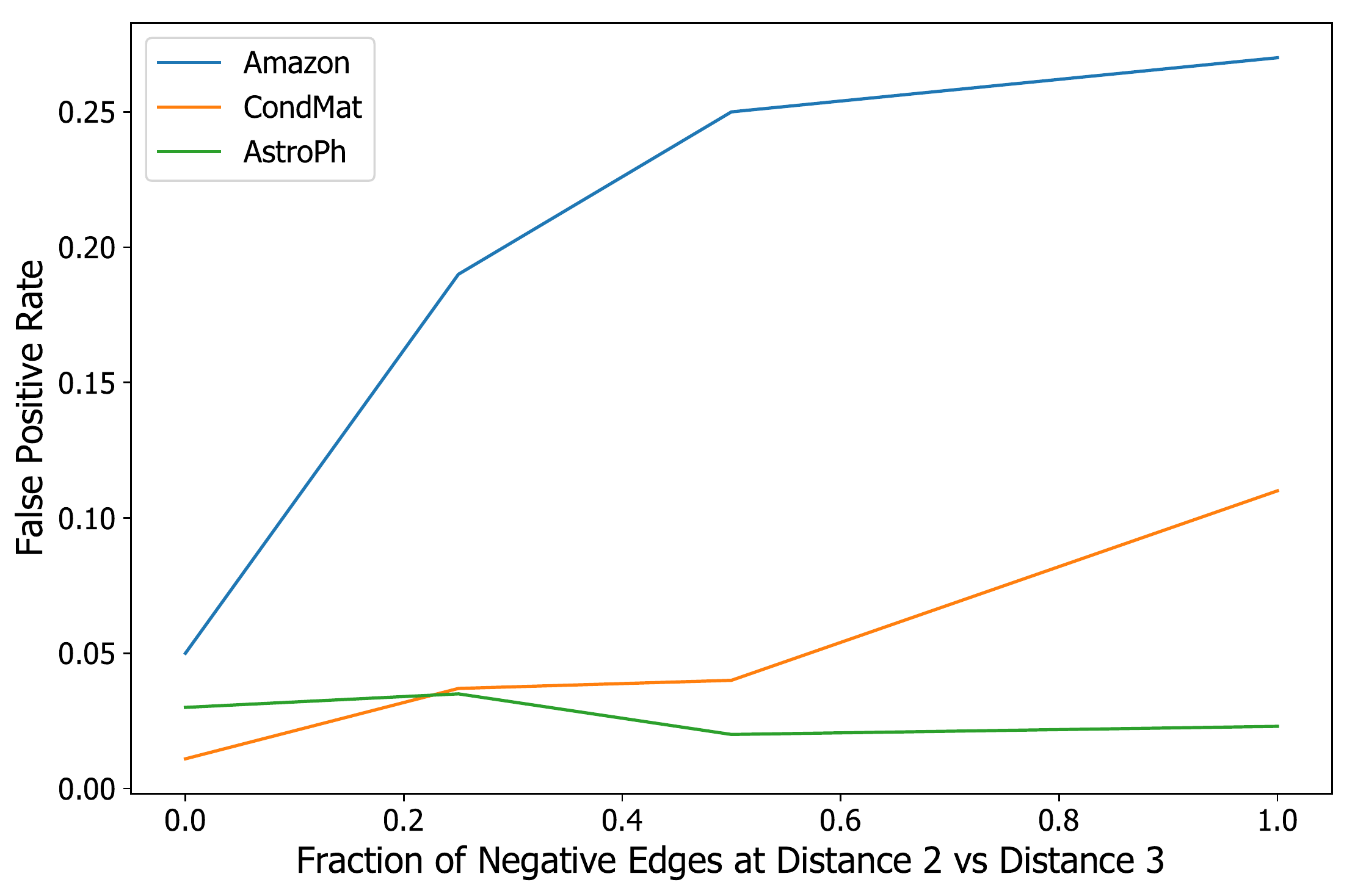}
\caption{False positive rate as a function of the fraction of negative examples at distance 2 (vs.\ distance 3).}
\label{fig:fpr-2to3}
\end{figure}

Figure~\ref{fig:acc-2to3} shows the classification accuracy for each setting.
For both Amazon and ContMat, the edges at distance 2 are harder to classify correctly, which produces a significant decline in the accuracy as we increase the fraction of edges at distance 2.
Conversely, the accuracy on AstroPh does not seem affected.
The same pattern can be seen in Figure~\ref{fig:fpr-2to3}, which reports the false positive rate.
The figure explains the cause of the decrease in accuracy: as we decrease the average distance of the negative examples, the classifier produces more false positives.
The higher the fraction of negative examples at distance 2, the higher the rate of misclassification for the negative class.
Again, AstroPh is the exception in our experiments.

We conclude that classification accuracy can be affected by the choice of the negative examples, and that the distance of negative samples should be carefully controlled.
In general, negative examples with smaller distances are harder to classify correctly, and they are more likely to be false positives.
However, the effect remains dataset-specific.

\subsection{Scalability of Motif Extraction}
In our final experiment, we show the time taken to extract 5-motifs on the Amazon graph, the largest of the graphs that we use. These times were measured while running Arabesque~\cite{teixeira2015arabesque} on 4 compute nodes with 8 cores and 256GB of memory each.

\begin{figure}
\centering
\includegraphics[width=\columnwidth]{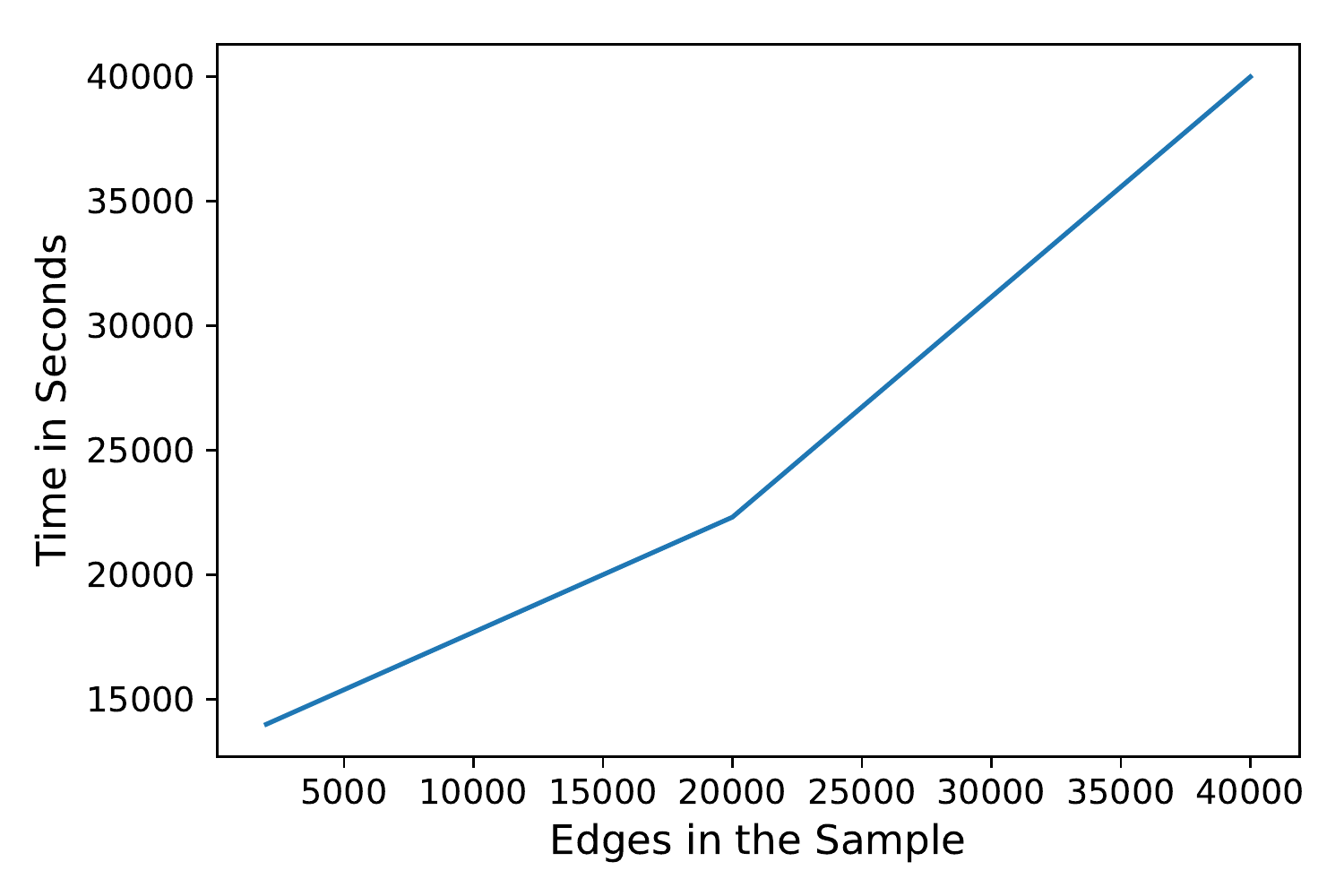}
\caption{Time required to extract 5-motifs (m5.*) for different sample sizes on the Amazon graph.}
\label{fig:time}
\end{figure}

We measure the extraction time while varying the number of sample edges for which to extract the motifs.
The results can be seen in Figure~\ref{fig:time} for samples of \num{2000}, \num{20000}, and \num{40000} edges.
The times taken are around, 4, 6, and 11 hours, respectively.
The lower efficiency for \num{2000} edges (500 edges/hour) is due to startup costs of the motif-counting framework.
For the larger samples of \num{20000} and \num{40000} edges, the startup costs are 
amortized over a larger number of edges leading to higher efficiency
(\num{3600} edges/hour for \num{40000} edges).
These results show that, with modern systems like Arabesque, extracting motifs up to size 5 is currently manageable for medium sized graphs.
Modern machine learning models are sometimes trained in days, whereas extraction times reported in Figure~\ref{fig:time} are in hours.

\section{Related work}
\label{sec:related}
There are two main branches of research that are relevant to the current work: graph motifs and link prediction.

\subsection{Graph Motifs}
Motifs are patterns of connectivity that occur significantly more frequently in the given graph than expected by chance alone~\citep{vazquez2004topological}.
Graph motifs have numerous applications, for example, they have been used to classify graphs into ``superfamilies''~\citep{schneider1991dominant}, and they have been used in combination with machine learning techniques to determine the most appropriate network model for a given real-world network~\citep{soutoglou2002coordination}.
\citet{palla2005uncovering} also show that 4-cliques (fully-connected motifs of 4 vertices) reveal community structure in word associations and protein-protein interaction graphs.

In several social media analysis studies~\cite{juszczyszyn2008local, juszczyszyn2011link, ugander2013subgraph}, graph motif detection and enumeration are used to characterize graph properties statistically.
The significance of motifs is typically assessed statistically by comparing the distribution of subgraphs in an observed graph with the one found in a randomized graph.
The randomization process is designed to alter the structure of the graph while preserving the number of nodes, edges, and the degree distribution~\citep{vazquez2004topological}.
One of the important reasons why graphs in the real world have more motif structure than the randomized version 
is that real-world graphs are constrained by particular types of growth rules, which in turn depend on the specific nature of the graph.
In this paper, we aim at leveraging this property to learn which specific motifs are predictive of link presence.


\subsection{Link Prediction}
Prior work on link prediction can generally be classified into three broad categories:
unsupervised methods, supervised methods, and feature learning methods.
Link prediction methods can also be orthogonally classified by the type of information they rely on: node properties or structural properties.
The two methodologies can always be combined when both types of information are available.

\spara{Unsupervised methods.}
In most unsupervised methods, a heuristic is chosen and used to rank node pairs in the graph,
with a higher rank indicating a higher likelihood of a link existing between the node pair~\citep{lu2010link, folino2012link}.
The heuristic is typically a similarity measure, and can be based on application-specific node attributes or on the graph topology.
Examples of node attributes used in unsupervised methods include user profiles in social networks~\citep{wang2015link}
and attributes of publication records in academic co-authorship graphs, such as the topics a researcher works on~\citep{chuan2017link} or the conferences they submit to~\citep{liang2017link}.
While node attributes can achieve a high degree of accuracy for link prediction, they are domain- and application-specific, and cannot be easily generalized.
In contrast, features based on graph topology are more general and directly applicable to any graph.

Topological features that are used in unsupervised link prediction are typically related to local (neighborhood) or global (path) properties of the graph.
Neighborhood-based features capture the intuition that a link is likely to exist between a pair of nodes if they have many common neighbors.
The simplest neighborhood-based feature is to count common neighbors (i.e., open triangles)~\cite{newman2001clustering}.
More advanced features include some form of regularization of the count, such as the Jaccard coefficient of the two sets of neighbors,
and the Adamic/Adar index~\cite{adamic2003friends}, which discounts the contribution of high-degree neighbors,
or preferential attachment~\cite{barabasi2009scale}, which gives a higher likelihood to links between high degree vertices.
Local similarity indices are easy to compute, and scale well to large graphs.

Conversely, path-based features look at the global graph structure.
A representative path-based feature is the classic Katz index~\cite{katz1953new}, which counts the number of paths between two nodes, giving a higher weight to shorter paths.
Other methods such as hitting time, commute time, and rooted PageRank use random walks on the graph to derive the similarity of two nodes.
Global similarity indices typically provide better predictions than local indices, but are more expensive to compute, especially in large graphs.
For a detailed survey of unsupervised link prediction methods, see the works by~\citet{getoor2005link} and~\citet{lichtenwalter2010new}.

Several studies indicate that unsupervised methods are fundamentally unable to cope with dynamics, imbalance, and other complexities of real-world graphs~\cite{al2006link,lichtenwalter2010new}.
However, similarity indices can easily be used by supervised methods as features for a machine learning model.

\spara{Supervised methods.}
In supervised methods, link prediction is usually cast as a binary classification problem.
The target class label indicates the presence or absence of a link between a node pair.
The predictor features are metrics computed from the graph structure or node attributes which describe the given pair.
This approach was first introduced by \citet{liben2007link}, who use graph topological features to study a co-authorship network.
The supervised approach is also used in several later works~\cite{sa2010supervised, leskovec2010predicting, al2006link}.

A key challenge for supervised link prediction is designing an effective set of features for the task.
Some works use simple topological features such as the number of common neighbors and the Adamic/Adar index~\cite{fire2011link}, while others use more  complex features~\cite{cukierski2011graph}.
For detailed surveys on supervised link prediction methods, please refer to~\cite{al2006link,liben2007link,gao2015link}.

 
Our method is also a supervised method based on graph topology.
Specifically, we train a classifier on our proposed motif-based features.
Our experiments show that this classifier outperforms one trained on a combination of the graph topology features used in the prior work mentioned above.
Our proposed  prediction method aims to improve on local measures by looking into higher-order structures among nodes (i.e., motifs).

Applying supervised methods to link prediction requires preparing a classification dataset for training the model and testing its performance.
To ensure that the model will not over-fit to a specific set of samples, a set of existing 
links
must be hidden from the model.
If we can observe the graph as it evolves over time, then one approach is to take snapshots of the graph at different times.
Edges present in the graph in a snapshot are used as the training set, while edges present in the graph at a later point form the test set. This approach is widely adopted by several studies~\cite{liben2007link}. 
If the graph evolution cannot be observed, the edges are usually sampled uniformly at random among the available ones.
In either case, a set of negative examples (pairs of unconnected nodes) needs to be sampled from (a snapshot of) the graph.
Our method for constructing the classification dataset has some unique features compared to prior work, 
as we discussed
in Section~\ref{sec:method}.

\mpara{Feature Learning.}
The most sophisticated approach to link prediction in the literature is to allow the model to learn by itself which (latent) features are important for the link prediction task.
Feature learning methods such as matrix factorization or deep learning
depend on the examination of graph topology and structural features to compute score functions based on pairwise similarity metrics.
 
Approaches based on matrix factorization model the graph as an $N \times N$ matrix in which $N$ represents the number of nodes, and then predict a link by using matrix decomposition.
For example, \citet{menon2011link} consider link prediction as a matrix completion problem and solve it using a non-negative matrix factorization (NMF) method.
The basic idea is to let the model learn latent features from the topological structure of a partially observed graph, and then use the model to approximate the unobserved part of the graph via matrix completion.
However, NMF performs best when combining the latent features with features based on node attributes.
In our work, we use only topological features, so we compare our method to NMF using the latent (topological) features, and show that our method significantly outperforms NMF.

Deep learning is another very popular form of feature learning.
In particular, graph convolutional networks (GCNs) have recently emerged as a powerful tool for representation learning on graphs~\cite{kipf2016semi}.
GCNs have also been successfully 
used
for link prediction~\cite{zhang2018link,schlichtkrull2018modeling}.
For example, SEAL~\cite{zhang2018link} is a framework which fits a graph neural network to small subgraphs around the example edges in the dataset.
By doing so, it learns latent features in the neighborhood structure of the graph which indicate the presence or absence of a link.
Therefore, it is very similar in spirit to our current work.
Nevertheless, we show that our motif-based features are able to achieve a much higher prediction accuracy than the one obtained by SEAL.

\section{Conclusion}
\label{sec:conc}

We presented a new approach for link prediction in undirected graphs that
relies on using the distribution of $k$-motifs that a pair of nodes appears in to predict whether a link exists
between these two nodes.
We use $k \in \{3, 4, 5\}$, which distinguishes our approach from prior approaches that rely on topological features:
prior approaches are based on common neighbors, while our approach is based on higher-order analysis of the topology.
Our approach treats the link prediction problem as a classification problem, so building the classification dataset is a key step.
We pointed out two issues related to this step that were not adequately addressed by prior work.
First, it is important to treat positive and negative example edges in the same way.
Prior approaches achieve this by removing positive example edges from the graph, and we showed that an alternative (and better) way is to insert negative example edges in the graph.
Second, when sampling pairs of nodes to find negative example edges, the shortest-path distance between the sampled nodes affects prediction accuracy, with shorter distances increasing the difficulty of the problem.
Thus, it is important to control 
this parameter when building the classification dataset (in our case we used a 50/50 split between distance 2 and 3).
Given a classification dataset constructed with these two issues properly addressed, we showed that it is possible, by using off-the-shelf classifiers, to achieve substantial improvement in prediction accuracy compared to prior methods based on topological features or feature learning.

For future work, it would be interesting to explore the use of graphlet-based features instead of motif-based features.
Graphlets are induced subgraphs, whereas the motifs that we currently use are partial non-induced subgraphs, and it
would be interesting to see if graphlets can provide accuracy gains if used alone or in combination with motifs.
Another direction for future work is exploring the temporal aspects of graph evolution.
Link prediction is ultimately a temporal problem, where we predict links in the future based on a snapshot of the graph in the present.
If every edge has a creation timestamp, temporal motifs can be defined by putting a label on each edge which indicates the relative order of its appearance in the motif.
These temporal motif features would capture the graph evolution at a fine resolution, and could perhaps increase the prediction accuracy.

\bibliographystyle{nourlabbrvnat}
\bibliography{references}

\balance




\end{document}